# Three-Dimensional Granular Flow Simulation Using Graph Neural Network-Based Learned Simulator

Yongjin Choi, M.S., S.M.ASCE[1]; Krishna Kumar, Ph.D., Aff. M. ASCE[2].

[1]Department of Civil, Architectural and Environmental Engineering, The University of Texas at Austin, TX, E-mail: yj.choi@utexs.edu
[2]Department of Civil, Architectural and Environmental Engineering, The University of Texas at Austin, TX, E-mail: krishnak@utexas.edu

**ABSTRACT**

Reliable evaluations of geotechnical hazards like landslides and debris flow require accurate simulation of granular flow dynamics. Traditional numerical methods can simulate the complex behaviors of such flows that involve solid-like to fluid-like transitions, but they are computationally intractable when simulating large-scale systems. Surrogate models based on statistical or machine learning methods are a viable alternative, but they are typically empirical and rely on a confined set of parameters in evaluating associated risks. Conventional machine learning models require an unreasonably large amount of training data for building generalizable surrogate models due to their permutation-dependent learning. To address these issues, we employ a graph neural network (GNN), a novel deep learning technique to develop a GNN-based simulator (GNS) for granular flows. Graphs represent the state of granular flows and interactions, like the exchange of energy and momentum between grains, and GNN learns the local interaction law. GNS takes the current state of the granular flow and estimates the next state using Euler explicit integration. We train GNS on a limited set of granular flow trajectories and evaluate its performance in a three-dimensional granular column collapse domain. GNS successfully reproduces the overall behaviors of column collapses with various aspect ratios that were not encountered during training. The computation speed of GNS outperforms high-fidelity numerical simulators by 300 times.

**INTRODUCTION**

Landslides entail massive granular flows and cause significant damage to civil infrastructures. Precise modeling runout caused by granular flows is critical to understand the impact of landslides. Numerical methods, such as DEM and MPM (Kumar et al., 2017; Mast et al., 2014; Utili et al., 2015; Yerro et al., 2019; Zenit, 2005), are often employed to assess landslide runouts. However, these methods are computationally expensive for simulating large-scale problems, hindering multiple full-scale simulations for a comprehensive evaluation of runout hazard scenarios. Similarly, a back analysis to estimate material parameters requires a broad parametric sweep involving hundreds to thousands of simulations. However, current state-of-the-art numerical methods are restricted to, at most, a few full-scale simulations, limiting our ability in scenario testing or back analysis.

To circumvent the computational burden, researchers build surrogate models based on statistical or machine learning approaches. These models typically use "end-to-end" mapping between landslide risks with influencing factors (Durante and Rathje, 2021; Gao et al., 2021; Ju et al., 2022; Zeng et al., 2021). Despite the success in associating granular flow runout with statistical or data-driven methods, these techniques do not directly account for the granular flow dynamics, the fundamental physics that governs flow behavior. Consequently, the absence of this physics consideration restricts these statistical models from extrapolating to different boundary conditions or geometries beyond the data on which they are built.

Establishing a surrogate model that can replicate granular flow dynamics is a challenging problem. It should capture a wide range of behavior changes—non-linear, static, collisional, and frictional dissipation regimes involved in granular flows (Soga et al., 2016). Learning to simulate these complex behaviors requires the surrogate model to understand fundamental interactions between neighbors. However, traditional machine learning techniques, such as multi-layer perceptron (MLP) or convolutional neural networks (CNN), face challenges in learning these behaviors. Since MLPs are permutation-dependent learning, meaning that their outputs are always associated with the order of inputs, they require unreasonably large training data to map all the possible permutations of the grain arrangements and the interactions involved in granular flows (Battaglia et al., 2018). CNN is restricted to learning mesh-based systems causing challenges when it comes to learning the dynamically changing neighbors.

To overcome these limitations, we use graph neural networks (GNNs), a state-of-the-art machine learning model (Battaglia et al., 2018; Battaglia et al., 2016; Sanchez-Gonzalez et al., 2020), to learn the interaction law involved in granular flows. We develop a GNN-based Simulator (GNS) that uses graphs to represent the state of interacting granular flow and learns the fundamental interaction based on GNN. We train the GNS on the limited number of granular flow trajectories generated from the material point method (MPM). The performance of GNS is tested using the granular column collapse experiment in a three-dimensional domain, which quantifies overall large-scale runout dynamics. GNS successfully predicts the overall runout dynamics of column collapse in a different boundary condition and initial geometry beyond the data being trained.

## METHOD

**Graph Neural Network-based Simulator (GNS).** Graphs $G = (V, E)$ are data that efficiently describe interactions between objects (Figure 1a). Graphs consist of vertices $v_i \in V$ representing objects and edges $e_{i,j} \in E$ connecting a pair of vertices ($v_i$ and $v_j$) representing the interactions between the objects with their direction represented with the arrows. Graph neural networks (GNNs) (Figure 1) take the graph as an input (Figure 1a), conduct the message passing, and return the updated graph $G' = (V', E')$ (Figure 1b) with the updated vertices $v'_i \in V'$ and edges ($e'_{i,j} \in E'$). Message passing is the fundamental operation of GNNs that models the information exchange among the vertices through the edges using neural networks.

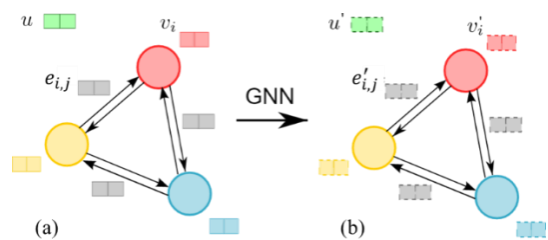

**Figure 1. Graph and graph neural network (GNN) (Modified from (Kumar and Vantassel, 2022))**

GNN-based simulator (GNS) proposed by (Sanchez-Gonzalez et al., 2020) predicts the next state of the flow $x_i^{t+1} \in X_{t+1}$ at timestep $t+1$ based on that of the current step $x_i^t \in X_t$ ($GNS: X_t \rightarrow X_{t+1}$) (Figure 2a). $x_i^t$ includes the information about position, velocities for five previous timesteps, distance to boundaries, and material properties of the $i$th material point at timestep $t$. GNS consists of dynamics approximator and updater. The dynamics approximator (Figure 2b) predicts the dynamics of material points $y_i^t \in Y_t$ using three stages of operations: encode-process-decode. In the encode stage, $X_t$ is encoded into a latent graph $G = (V, E)$ to represent the state of interacting material points. We use MLP as the encoder. In the process stage, multiple stacks of GNN layers update $G = (V, E)$ to $G' = (V', E')$ through message passing. This stage models the interaction between grains such as energy or momentum exchange, which enables GNS to learn the interaction law. In the decode stage, the dynamics of material points $y_i^t \in Y_t$ is extracted from $G'$. We use MLP as the decoder. In the updater (Figure 2a) takes the $y_i^t \in Y_t$ and updates the current state $x_i^t \in X_t$ to next state $x_i^{t+1} \in X_{t+1}$. This process is similar to explicit Euler integration in numerical differential equation solver; therefore, we can consider $y_i^t \in Y_t$ as time derivatives such as acceleration. GNS predicts entire trajectory of granular flows $X_0, X_1, \ldots, X_k$, where $X_0$ is the initial state and $X_k$ is the state at timestep $k$, by updating $X_t$ to $X_{t+1}$. For more details about GNNs, message passing, and GNS, we refer readers to Choi and Kumar (2023).

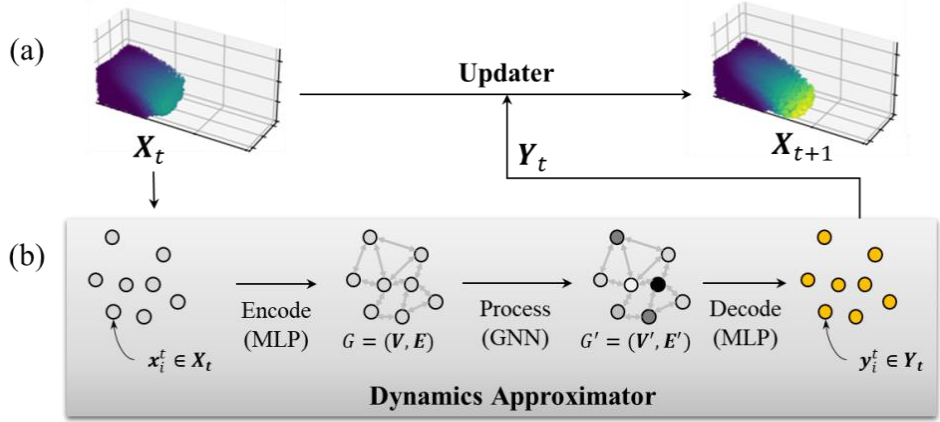

**Figure 2. Graph Neural Network-based Simulator (GNS).**

**Training.** The encode and decode stage both use two layers of 128-dimensional MLP layers with learnable parameter sets $\theta_\epsilon$ and $\theta_\delta$ where $\theta_\epsilon$ is for encode MLP, and $\theta_\epsilon$ is for decode MLP. The process stage consists of 10 stacks of graph neural networks with a learnable parameter set $\theta_\eta$. Given the current state $X_t$, we train these parameter sets ($\theta_\epsilon$, $\theta_\delta$, and $\theta_\eta$) in GNS to minimize the mean squared error between the current ground truth acceleration $A_t$ and the predicted dynamics $Y_t$ of material points at $t$.

To create training examples (i.e., sets of $X_t$ and $A_t$), we use the material point method (MPM) based on CB-Geo MPM code (Kumar et al., 2019). The training examples are sampled from the trajectories of 106 granular flow simulations generated from MPM. The simulation starts with a cube-shaped mass with varying size and initial velocities, and it drops under gravity in the 1.0×1.0×1.0 m cube domain (Figure 3). We restrict the shape of the mass to a cube to test the generalizability of the trained GNS to unseen geometries that show different flow dynamics. The computation is conducted using a cube-shaped mesh with a length of 0.0833 m including 16 material points in it. Simulation $dt$ is 1e-6 and the result is saved for every 2500 timesteps. The visual examples of the simulation results are sampled in the figure below in four chronological orders. Note that our available trajectory data for making training data (106 trajectories) is far less than the numbers typically required to train GNS (1000 trajectories) (Sanchez-Gonzalez et al., 2020).

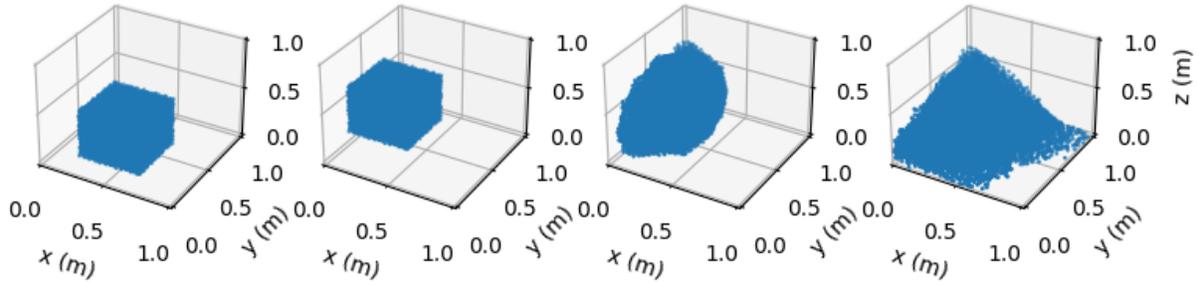

**Figure 3. An example of the trajectory of training data. The cube-shaped mass drops and flows under gravity with a random initial velocity and position.**

**Evaluation.** We evaluate the performance of GNS using the granular column collapse experiment: a cuboid-shaped granular column is placed on a flat surface of one end of the boundary, and it is allowed to collapse by gravity. The collapse shows different flow dynamics depending on the initial aspect ratio ($a = H_0/L_0$ where $H_0$ and $L_0$ is the initial height and length of the column before collapse) of the column (Lube et al., 2005). When the column has a small aspect ratio ($a \lesssim 1.7$) (i.e., short column), only the flank of the column is mobilized during the failure leaving the majority of the soil mass under the failure surface static. On the other hand, when the column has a large aspect ratio ($a \gtrsim 1.7$), the collapse starts with a free-fall-like drop at the upper part of the granular mass, mobilizing large potential energy. The potential energy builds large kinetic energy causing longer a runout compared to the short column. As noted earlier, we only train GNS on granular flows with the initial aspect ratio of 1.0 in 1.0×1.0×1.0 m cube domain, but we test the predictive performance of GNS for runout dynamics on a different simulation domain with 1.5×0.5×0.7 m, with the columns with different aspect ratios, including short ($a = 0.8$) and tall column ($a = 2.0$), which are unseen during training.

## Result and Discussion

**Short Column.** Figure 4 shows flow evolution with normalized time ($t/\tau_c$) for the short column with $a = 0.8$ from GNS and MPM. MPM is our baseline high-fidelity simulator. The column includes about 10K material points. The notation $t$ represents physical time, and $\tau_c \left(= \sqrt{\frac{H_0}{g}}\right)$ represents the critical time representing time required for the flow to fully mobilize where $g$ is the gravitational acceleration. Each row of the figure is the flow at the initial state before failure initiates, at normalized time 1.0, 2,5, and at the final time when the flow reaches static equilibrium.

The geometric evolution of the collapse shows the following three stages. (1) From $t/\tau_c = 0$ to 1.0 (Figure 4a), the flow starts with the collapse at the flank of the column and reaches full mobilization. (2) From $t/\tau_c = 1.0$ to 2.5 (Figure 4b) major flow spreading occurs. (3) After $t/\tau_c = 2.5$, the flow deaccelerates due to frictions among the boundaries and materials, and finally reaches static equilibrium (Figure 4c). This geometric evolution predicted by GNS shows overall agreement with MPM well replicating typical flow dynamics of short columns. However, GNS shows a larger volume of flank mobilization compared to MPM, leaving a smaller plateau area at the top of the static part (Figure 3d).

We also quantitatively investigate the GNS prediction. **Error! Reference source not found.** shows the normalized runout (($L_t - L_0/L_0$)) and normalized height ($H_t/L_0$) evolution with normalized time ($t/\tau_c$). $L_t$ is the distance from the left boundary to the front end of the flow, and $H_t$ is the distance from the base to the highest part of the column, at timestep $t$. The normalized runout evolution also shows three stages. (1) From $t/\tau_c = 0$ to 1.0, runout is slowly accelerated as the flank of the column collapses. (2) From $t/\tau_c = 1.0$ to 2.5, the major runout occurs. (3) After $t/\tau_c = 2.5$, runout shows deacceleration. The general runout trend of GNS follows the result from MPM, but the smaller runout is observed as time evolves with the error of 12% at the final time when flow ceases. For both GNS and MPM, the height remains steady since only the flank of the column collapses leaving a static plateau.

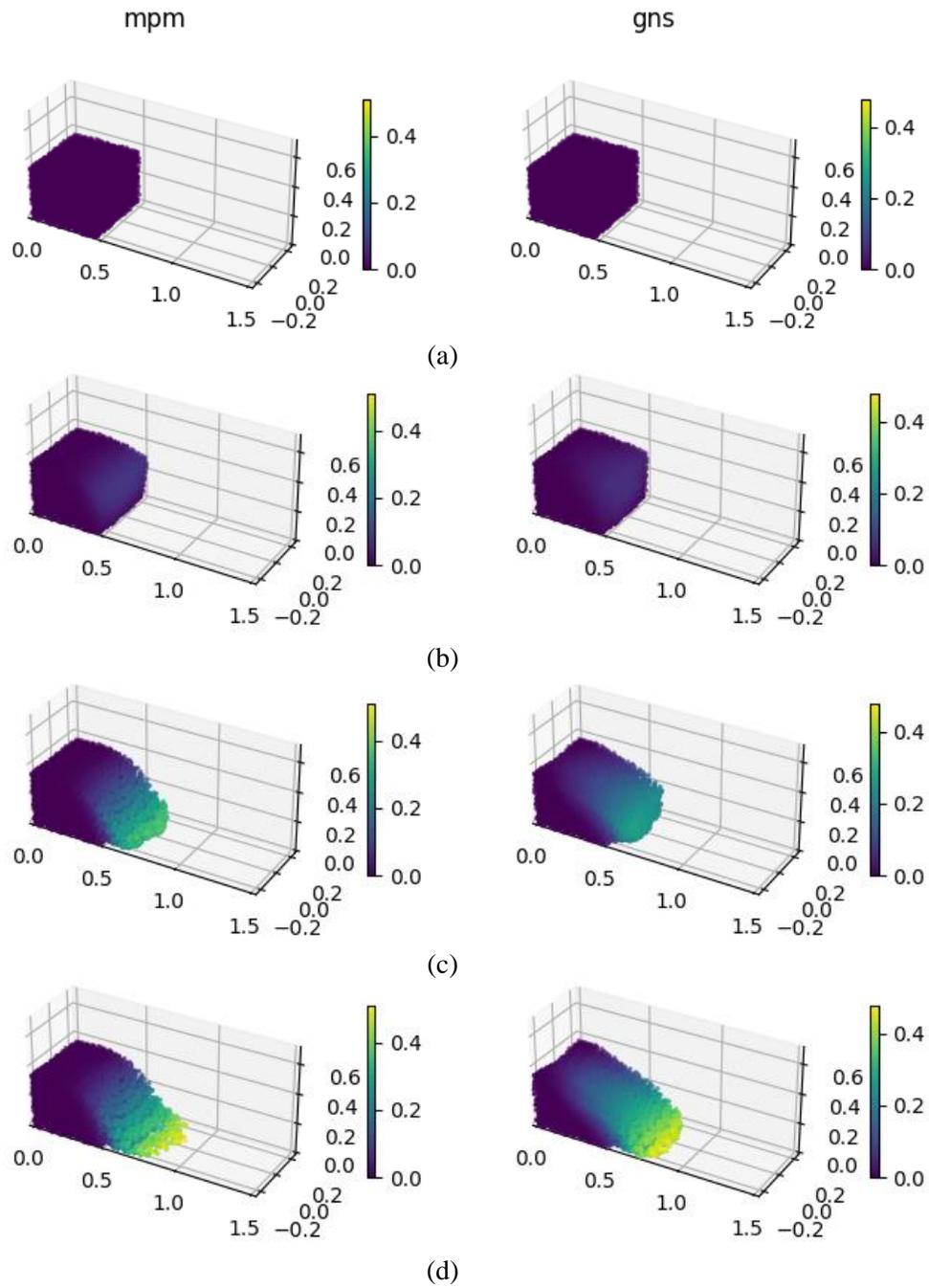

**Figure 4. Evolution of flow with normalized time for GNS and MPM for the short column with $a = 0.8$: (a) initial state, (b) $\frac{t}{\tau_c} = 1.0$, (c) $\frac{t}{\tau_c} = 2.5$, (d) final state. Units are in m. The color represents the magnitude of the displacement.**

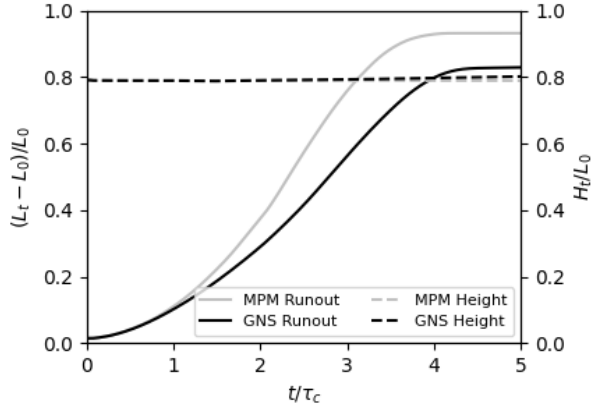

**Figure 5. Normalized runout $((L_t - L_0)/L_0)$ and normalized height $(H_t/L_0)$ evolution with normalized time $(t/\tau_c)$ for the short column with $a = 0.8$.**

**Tall Column.** Figure 6 shows flow evolution with normalized time $(t/\tau_c)$ for GNS and MPM for the tall column with $a = 2.0$. The column includes about 10K material points. Similar to the short column collapse, three collapse stages are observed: initial flow mobilization (from initial state to $t/\tau_c = 1.0$), major runout (from $t/\tau_c = 1.0$ to 2.5), deceleration (after $t/\tau_c = 2.5$) to reach static equilibrium. However, compared to the short column, a relatively larger runout is observed with a greater collapsing mass volume. This is because the larger potential energy release due to the free-fall-like drop of the mass during the initial stage and the subsequent kinetic energy build-up cause a substantial horizontal acceleration. In general, the geometric evolution of GNS for the tall column well replicates the result from MPM, but we observe a larger mobilized mass in GNS (**Error! Reference source not found.**d) leading to a lower final height than MPM.

Figure 7 shows the normalized runout $((L_t - L_0/L_0))$ and normalized height $(H_t/L_0))$ evolution with normalized time. The GNS predicts a larger height settlement than MPM due to the larger collapsing mass during the flow. For the runout, GNS predictions exhibit a similar trend with MPM maintaining slightly smaller values compared to MPM. The GNS accurately estimates the final normalized runout with an error of 2%.

**Computation Efficiency.** One of our objectives is to accomplish faster computation time than the high-fidelity numerical method. Here we compare the computation time between GNS and MPM to simulate the short and tall column collapse shown in **Error! Reference source not found.** and Figure 7. MPM was run on 56 cores of Intel Cascade 495 Lake processors in parallel, and GNS was run on a single RTX GPU on TACC Frontera systems. MPM requires approximately 6 hours to finish the computation, while GNS requires 80 s to compute the trajectory for 380 timesteps, which achieves 300× speed-up.

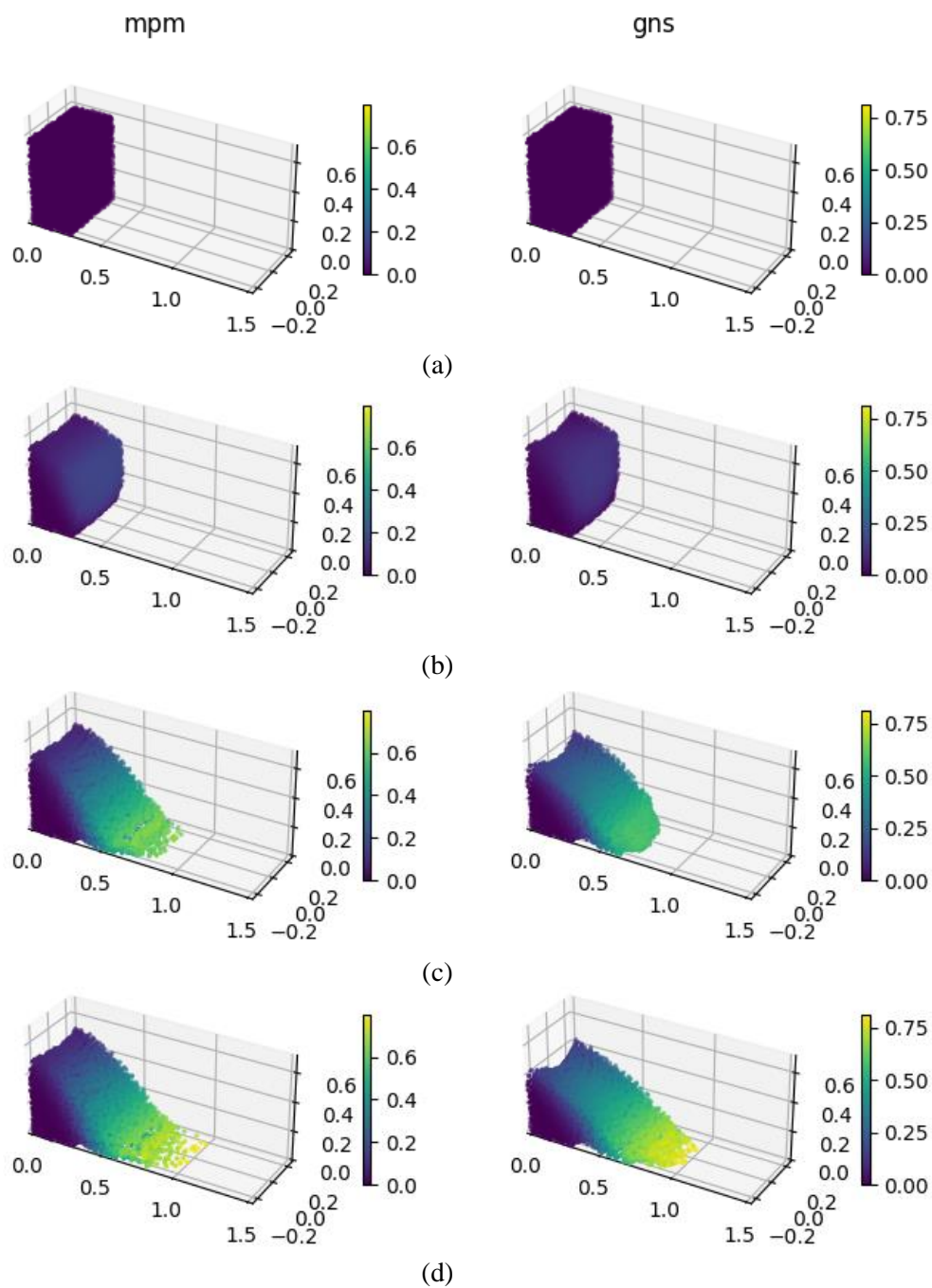

**Figure 6. Evolution of flow with normalized time for GNS and MPM for the tall column with a = 2.0: (a) initial state, (b) $\frac{t}{\tau_c} = 1.0$, (c) $\frac{t}{\tau_c} = 2.5$, (d) final state. Units are in $m$. The color represents the magnitude of the displacement.**

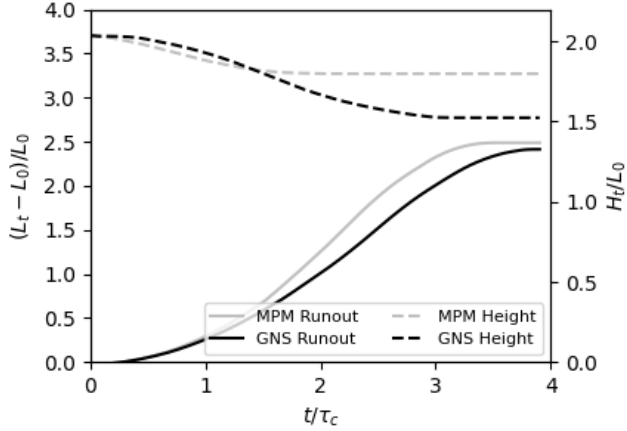

**Figure 7.** Normalized runout $((L_t - L_0)/L_0)$ and normalized height $(H_t/L_0)$ evolution with normalized time $(t/\tau_c)$ for the tall column with $a = 2.0$.

## DISCUSSION

Our GNS trained on 106 granular flow trajectory data is able to generalize well to replicate the overall flow dynamics of granular column collapse (Figure 4 and Figure 6), although the boundary and the initial geometry of the granular mass diverge from the training data. However, we observe some differences in the quantitative values for runout and height evolution compared to MPM (**Error! Reference source not found.** and Figure 7). Specifically, GNS shows a larger height settlement for the tall column case, and it also shows a slightly shorter runout for the short column. This difference can be mainly attributed to the limited training data—as mentioned earlier, the number of trajectory data available is 106 while usually 1000 trajectories are used to train GNS based on Sanchez-Gonzalez et al. (2020). Nevertheless, the computational efficiency of the GNS makes it valuable for use in preliminary analysis before conducting full-scale simulation using high-fidelity numerical methods. For example, Kumar et al. (2022) applied GNS as an oracle for in-situ visualization for granular flows to identify critical regions before running the large-scale MPM simulation, utilizing the computation efficiency of the GNS.

## CONCLUSION

Traditional numerical methods, such as MPM and DEM, are computationally intractable in large-scale granular flow simulations, which hinders multiple scenario testing and parameter calibration. Typical statistical or conventional machine learning-based surrogate models mapping the risks associated with granular flows and affecting factors do not explicitly consider the underlying physics, limiting their effectiveness and generalizability. To overcome these challenges, we use graph neural networks (GNNs), a state-of-the-art deep learning model, to develop a learned simulator, GNS. The physical state of interacting granular flows is represented by graphs, and GNN processes the graphs using message passing which learns to model the complex interaction between grains. Graph representation and message passing enable accurate learning to predict the granular flow dynamics across different conditions, even those unseen during training. We evaluate the performance of GNS on the three-dimensional granular column collapse experiment. GNS can be generalized to different flow dynamics stemming from varying initial aspect ratios in a different simulation domain not trained during the training process. In addition, GNS exhibits an outstanding computation speed that is 300 times faster than the parallelized CPU version of MPM, while maintaining reasonable accuracy. The computational efficiency and generalizability of GNS will make it a valuable tool for efficiently assessing runout hazards in a wide range of scenarios.